\documentclass[twocolumn,showpacs,preprintnumbers,amsmath,amssymb,prc]{revtex4-1}
\usepackage{epsfig}

\usepackage{mathptmx, courier, pifont}
\usepackage[scaled=0.92]{helvet}
\usepackage[T1]{fontenc}
\usepackage{textcomp}

\begin{document}

\author{Ji-Wei Cui$^{1}$, Xian-Rong Zhou$^{1}$\footnote{Corresponding author: xrzhou@xmu.edu.cn}, Fang-Qi
Chen$^{2}$, Yang Sun$^{2}$, Cheng-Li Wu$^{1}$, Zao-Chun Gao$^{3}$}

\address{$^{1}$Department of Physics and Institute of Theoretical Physics and Astrophysics, \\
 Xiamen University, Xiamen 361005,People's Republic of China\\
$^{2}$Department of Physics and Institute of nuclear, particle, astronomy and cosmology, \\
Shanghai Jiao Tong University, Shanghai 200240, People's Republic of China\\
$^{3}$China Institute of Atomic Energy, P.O. Box 275(18) Beijing 102413, People's Republic of China }

\title{Description of collective and quasiparticle
excitations in deformed actinide nuclei:\\ The first application of
the Heavy Shell Model}

\date{\today}

\begin{abstract}
The Heavy Shell Model (HSM) (Y. Sun and C.-L. Wu, Phys. Rev. C {\bf
68}, 024315 (2003)) was proposed to take the advantages of two
existing models, the projected shell model (PSM) and the Fermion
Dynamical Symmetry Model (FDSM). To construct HSM, one extends the
PSM by adding collective $D$-pairs into the intrinsic basis. The HSM
is expected to describe simultaneously low-lying collective and
quasi-particle excitations in deformed nuclei, and still keeps the
model space tractable even for the heaviest systems. As the first
numerical realization of the HSM, we study systematically the band
structures for some deformed actinide nuclei, with a model space
including up to 4-quasiparticle and 1-$D$-pair configurations. The
calculated energy levels for the ground-state bands, the collective
bands such as $\beta$- and $\gamma$-bands, and some quasiparticle
bands agree well with known experimental data. Some low-lying
quasiparticle bands are predicted, awaiting experimental
confirmation.
\end{abstract}

\pacs{ 21.10.Re, 21.60.Cs, 23.20.Lv}

\maketitle

\section{INTRODUCTION}

The interplay between collective motion and quasi-particle
excitations has been a long-standing topic in nuclear structure
physics. The nuclear shell model is the most fundamental method
that treats nuclear systems fully quantum mechanically in terms of
nucleons. However, it is difficult for the conventional shell
model based on a spherical basis to study heavy, deformed nuclei
because of the problem of huge dimensionality. Even with the
today's computer power and novel diagonalization algorithms, a
full shell model calculation for an arbitrarily large system seems
to be impossible. To overcome the dimensionality problem, one
needs to seek judicious truncation schemes and use more efficient
shell-model bases. In the literatures, the Projected Shell Model
(PSM)~\cite{Har95} and the Fermion Dynamical Symmetry Model
(FDSM)~\cite{Wu94} are two such examples. Both of them are based
on the shell model concept, but are constructed according to
different truncation schemes, thus emphasizing different physical
aspects.

In the PSM, shell model diagonalization is carried out in the
projected deformed basis constructed by choosing a few
quasiparticle (qp) orbitals near the Fermi surfaces and performing
angular-momentum and particle-number projection on the qp
configurations \cite{Har95}. In this way, the PSM is able to
describe low-lying rotational bands built upon qp excitations. It
has been successful for the PSM to study the rotational states in
heavy~\cite{Sun04} and superheavy nuclei~\cite{Falih09}, as well
as the states of super-deformation \cite{Sun95,Sun97}. Moreover,
it has been shown that comparing with the large-scale shell model
calculations~\cite{Cau95}, the PSM can achieve a similar accuracy
in describing the deformed $^{48}$Cr~\cite{Har09} and the
superdeformed $^{36}$Ar~\cite{Lon01}. However, the original
version of the PSM was not designed to treat collective
vibrational states such as $\beta$- and $\gamma$-vibrations. The
lack of ingredients for collective excitations in the PSM makes it
difficult to produce these low-lying collective bands, and also
limits its applications only to well deformed nuclei. To release
the restriction of axial symmetry in the deformed basis, the
Triaxial Projected Shell Model (TPSM) was introduced
\cite{Sheikh99}. A more recent example of the TPSM application is
to describe the $\gamma$-vibrational bands in some Er
isotopes~\cite{TPSM}.

On the other hand, the Interacting Boson Model (IBM)~\cite{Iac87} is
a successful model for the description of low-lying collective
states. In this model, the coherent $S$- and $D$-pairs are assumed to be
the building blocks of the low-lying collective states, and are
approximated as ${\it s}$ and ${\it d}$ bosons. It has been shown
that an axially symmetric rotor possesses the SU(3) symmetry, while
a $\gamma$-soft rotor possesses the SO(6) symmetry. The $\beta$- and
$\gamma$-vibrations including the scissors mode vibration in
deformed nuclei can be classified as different SU(3) or SO(6)
irreducible representations~\cite{Iac84}. Since nucleons are
fermions, the later developed FDSM directly uses coherent nucleon $S$-
and $D$-pairs without a boson approximation. The FDSM actually uses a
symmetry-dedicated shell model truncation scheme to treat nuclear
collective excitations. It has been shown that the FDSM can well
describe the low-lying collective states from the spherical to the
well-deformed region~\cite{Wu94}. However, the FDSM has difficulties
in describing single-particle excitations, because once the unpaired
single-particle degrees of freedom are opened up, the dimension of
the model space will go up quickly just like the conventional
spherical shell model. Moreover, the FDSM is a one-major-shell shell
model.

It is clear that both the PSM and the FDSM follow the shell model
philosophy, but they employ different truncation schemes, thus
describing different excitation modes. The PSM emphasizes qp
excitations, while the FDSM emphasizes low-lying collective
excitations. Experimentally, it is often the case that
quasi-particle and collective excitations coexist in the low-lying
nuclear spectrum. It is therefore desired to combine the advantages
of these two models to form a new shell model for heavy nuclei,
which can describe both qp excitations and low-lying collective
excitations simultaneously. The combination of the two models
becomes possible through the recognition~\cite{Sun98,Sun02} that the
PSM calculations exhibit, up to high angular momenta and excitation
energies, a remarkable one-to-one correspondence with the analytical
SU(3) spectrum of the FDSM. Motivated by this finding, it was
suggested in Ref. \cite{Sun03} that it is possible to treat
collective and qp excitations in a common multi-shell shell-model
framework. One way to realize the idea is to extend the PSM by
adding the coherent $D$-pairs into the intrinsic basis, since it is
evident from the FDSM that it is the coherent $D$-pairs that are
responsible for the collective excitations. With this extension, the
PSM ($i. e.$ the HSM) may become a more general multi-major-shell
shell model, useful not only for well-deformed nuclei, but hopefully
also for transitional ones (see discussions in Ref. \cite{Sun03}).

The key question for implementing the HSM is how to construct the
$D$-pairs in the PSM model space, which usually involves three
major shells for both neutrons and protons. In Ref.~\cite{Sun08},
the $D_{0}$ ($D_{2}$)-pair was suggested to be the linear
combination of all the 2-qp states with $K^{\pi}=0^{+}$
$(K^{\pi}=2^{+})$ in the PSM multi-major-shell truncated space.
The structure amplitudes are obtained from the wavefunction of the
lowest 2-qp state after diagonalization. A testing calculation was
performed for the $\beta$-band in $^{172}$Yb, and it was found
that indeed, the collective nature of the $D_{0}$ configuration
can be well reproduced from the calculation~\cite{Sun08}. In
Ref.~\cite{Cui&Zhou}, it was shown that by including both qp and
D$_{0}$ configurations, the ground-state bands (g-bands) and
$\beta$-bands of four deformed nuclei, $^{230, 232}$Th and $^{232,
234}$U in the actinide region, are also well reproduced. In
addition, the calculated $B(E2)$ transition rates agree well with
the experimental data. The structure of the $D_{0}$-pair in the
calculation does show collectivity. It is indeed a strong mixture
of many 2-qp states. All these indicate that the suggested
construction \cite{Sun08} of $D_{0}$-pair is reasonable.

The above attempts may be regarded as an initial step of the
numeric realization of the HSM. However, in order to describe
$\gamma$-bands, one needs to add the $D_{2}$-pair into the PSM
basis. Furthermore, in order to have the so-called `2-phonon
states' one needs to consider 2-$D$-pair excitations. As suggested
by the FDSM, the 2-$D$-pair excitations have four different
excitation modes: $D_{0}D_{0}$, $D_{2}D_{0}$, $D_{2}D_{-2}$ and
$D_{2}D_{2}$, which will give rise to the following four
2-phonon-excitation bands: $\beta\beta$-band (${\it
n}_{\gamma}=0$, ${\it n}_{\beta}=2$, $K/2=0$), $\gamma\beta$-band
(${\it n}_{\gamma}=0$, ${\it n}_{\beta}=1$, $K/2=1$),
$\gamma\gamma(0^{+})$-band (${\it n}_{\gamma}=2$, ${\it
n}_{\beta}=0$, $K/2=0$), and $\gamma\gamma(4^{+})$-band (${\it
n}_{\gamma}=1$, ${\it n}_{\beta}=0$, $K/2=1$), respectively, where
${\it n}_{\beta}$, ${\it n}_{\gamma}$ and $K$ denote the quantum
numbers of $\beta$- and $\gamma$-phonons and the z component of
angular momentum. In Refs.~\cite{Zeng1,Zeng2}, the rotational
bands in the nuclei with Z=100 were investigated systematically by
using cranking shell model with the pairing correlations treated
by a particle-number conserving method. In the present paper, we
study systemically the band structure of both the low-lying qp and
collective excitations for the deformed actinide nuclei
$^{230,232}$Th, $^{232,234,236}$U and $^{240}$Pu by adding
1-$D$-pairs into the PSM basis.

This paper is organized as follows. A brief introduction of HSM is
given in Sec. II. In Sec. III, we discuss in detail the structure
of $D_{0}$ and $D_{2}$ pairs, the energy schemes, eigen-functions
and reduced $B(E2)$ transitions for the nuclei $^{230,232}$Th,
$^{232,234,236}$U and $^{240}$Pu, respectively. Finally, a
conclusion is drawn in Sec. IV.

\section{FORMULISM}

The HSM is an improved version of PSM including not only single
particle excitations but also collective excitations in the basis.
However, the PSM cannot use directly the $D$-pair defined in the FDSM,
since the two model spaces are very different. The structure of
$D$-pairs is suggested in Ref.~\cite{Sun08} as follows:
\begin{equation}
D^{\dag}_{0}=\sum_{\rho,\mu}f_{\rho\mu}^{K=0}[a^{\dag}_{\rho}a^{\dag}_{\mu}]^{K=0},
~~~~~~
D^{\dag}_{2}=\sum_{\rho,\mu}f_{\rho\mu}^{K=2}[a^{\dag}_{\rho}a^{\dag}_{\mu}]^{K=2}.
\label{betagamma}
\end{equation}
where $[a^{\dag}_{\rho}a^{\dag}_{\mu}]^K $ is the 2-qp creation
operator with $K=0, 2$. $\rho$ and $\mu$ are the state index of
the qp, and $f_{\rho\mu}^K $ is the structure amplitude, which are
determined by diagonalizing the Hamiltonian in the 2-qp basis with
given $K$. Having $D^{\dag}_{0}$ and $D^{\dag}_{2}$ determined,
the one $D$-pair excitation will give the first $\beta$- and
$\gamma$-band. They can be expressed as
\begin{equation}
\arrowvert\ I,M \rangle_\beta =\hat{P}^{I}_{M0}D^{\dag}_{0}
\arrowvert\Phi\rangle, ~~~~~~ \arrowvert\ I,M \rangle_\gamma
=\hat{P}^{I}_{M2}D^{\dag}_{2} \arrowvert\Phi\rangle, ~~~~~~
\label{Probetagamma}
\end{equation}
where $\arrowvert\Phi\rangle$ is the BCS vacuum and
\begin{equation}
\hat{P}^{I}_{MK}=\dfrac{2I+1}{8}\int
d\Omega\hat{D}^{I}_{MK}\hat{R}(\Omega)  \label{PIMK}
\end{equation}
is the angular momentum projection operator. In Eq.~\eqref{PIMK},
$D^{I}_{MK}$ is the matrix element of $D$-function and $\hat{R}$ is
the rotation operator with respect to the solid angle $\Omega$ that
is always denoted by three Eular angles ($\alpha$, $\beta$,
$\gamma$). In our calculation, the axial symmetry in the deformed
basis is assumed, so $D$-function reduces to $d$-function and $\Omega$
reduces to $\beta$. Finally, adding the collective excitations into
the PSM intrinsic basis, the HSM intrinsic basis is given. For
even-even nuclei they are
\begin{equation}
\begin{array}{c}
\{\arrowvert\phi_\kappa\rangle\}=\{\arrowvert\Phi\rangle,
a^{\dag}_{\nu_i}a^{\dag}_{\nu_j}\arrowvert\Phi\rangle,
a^{\dag}_{\pi_k}a^{\dag}_{\pi_l}\arrowvert\Phi\rangle,\\
a^{\dag}_{\nu_i}a^{\dag}_{\nu_j}a^{\dag}_{\pi_k}a^{\dag}_{\pi_l}\arrowvert\Phi\rangle,
D^{\dag}_{0}\arrowvert\Phi\rangle,D^{\dag}_{2}\arrowvert\Phi\rangle
\}
\end{array}
\label{basis}
\end{equation}
where $a_{\nu_{i}}^\dag$ and $a_{\pi_{i}}^\dag$  are the qp creation
operators for neutrons and protons with $i$ as state index,
respectively.

The shell-model configuration space can then be constructed by the
projected basis, which is
\begin{equation}
\arrowvert
K,\kappa,IM\rangle=\hat{P}^{I}_{MK}\arrowvert\phi_{\kappa}\rangle,
\label{state}
\end{equation}
where $\arrowvert\phi_{\kappa}\rangle$ denotes the intrinsic basis
of HSM given in Eq.~\eqref{basis}. Then we can obtain the
eigen-energy $E^{\sigma}$ and the eigen-wavefunction
\begin{equation}
\arrowvert\Psi^{I,\sigma}_{M}\rangle=\sum_{K,\kappa}F^{I,\sigma}_{K,\kappa}\arrowvert K,\kappa,IM\rangle,
\label{eigenstate}
\end{equation}
where $\sigma$ denotes different eigen-states, by solving the
following eigenvalue equation:
\begin{equation}
\sum_{{K}^{'},{\kappa}^{'}}\left(\hat{H}^{I}_{{K\kappa},{K}^{'}{\kappa}^{'}}-E^{\sigma}\hat{N}^{I}_{{K\kappa},{K}^{'}{\kappa}^{'}}\right)
F^{I \sigma}_{K^{'}\kappa^{'}}=0, \label{eigeneq}
\end{equation}
where the Hamiltonian matrix element and the norm matrix element are
\begin{equation}
\hat{H}^{I}_{{K\kappa},{K}^{'}{\kappa}^{'}}=\langle\phi_{\kappa}\arrowvert\hat{H}\hat{P}^{I}_{K{K}^{'}}\arrowvert\phi_{{\kappa}^{'}}\rangle,
\label{Helemt}
\end{equation}
\begin{equation}
\hat{N}^{I}_{{K\kappa},{K}^{'}{\kappa}^{'}}=\langle\phi_{\kappa}\arrowvert\hat{P}^{I}_{K{K}^{'}}
\arrowvert\phi_{{\kappa}^{'}}\rangle.
\label{Nelemt}
\end{equation}
The effective interaction employed in the HSM is the same as that in
the PSM, which takes the form:
\begin{equation}
\begin{array}{c}
\hat{H}=\sum_{\xi=\nu,\pi}\hat{H}_{\xi}+\hat{H}_{\nu\pi},\hspace{0.5cm} \hat{H}_{\nu\pi}=-\chi_{\nu\pi}\hat{Q}_{2}^{\nu\dag}\hat{Q}_{2}^{\pi},\\
\hat{H}_{\xi}=\hat{H}^{\xi}_{0}-\frac{\chi_{\xi}}{2}\hat{Q}_{2}^{\xi\dag}\hat{Q}_{2}^{\xi}-G^{\xi}_{M}\hat{P}^{\xi\dag}\hat{P}^{\xi}-G^{\xi}_{Q}
\hat{P}_{2}^{\xi\dag}\hat{P}_{2}^{\xi}.\\
\end{array}
\label{Hamiltonian}
\end{equation}
The first term $\hat{H}_{0}^{\xi}$ in Eq.~\eqref{Hamiltonian} is the
spherical single-particle Hamiltonian. The second term is the
residual quadrupole-quadrupole interaction while the third and
fourth terms are the monopole-pairing and quadrupole-pairing
interactions, respectively. The strength of the
quadrupole-quadrupole force is determined by a self-consistent way
that would give the empirical deformation as predicted in the
variation calculation. The monopole-pairing strength is given as
follows
\begin{equation}
\begin{array}{l}
G^{n}_{M}=\left(19.3-0.08(N-Z)\right)/A,\\
G^{p}_{M}=\left(13.3+0.217(N-Z)\right)/A,
\end{array}
\label{Gstrenth}
\end{equation}
where `n' for neutrons and `p' for protons, respectively. The
monopole-pairing strength above is determined by reproducing the
experimental odd-even mass difference as Ref.~\cite{GMM79}. In the
current calculation it is multiplied by 0.87 in the cases of both
neutrons and protons. The quadrupole-pairing strength G$_{Q}$ is
proportional to G$_{M}$ and the proportional rate G$_{Q}$/G$_{M}$
is fixed to 0.14 in our calculation for $^{230,232}$Th, 0.13 for
$^{232,234,236}$U, and 0.12 for $^{240}$Pu. The parameters we
choose are slightly different from Ref.~\cite{Cui&Zhou} and
Ref.~\cite{Gao06, YS08} due to the different spaces used in our
present model. In FDSM, to produce the SU(3) symmetry, the
quadrupole-pairing strength is equal to the monopole-pairing
strength~\cite{Wu94}. The origin of the difference remains a very
interesting topic.

In Eq.~\eqref{Hamiltonian}, the one-body operator takes the
following form:
\begin{equation}
\begin{array}{l}
\hat{Q}_{\mu}=\sum_{\alpha,\alpha^{'}}Q_{\mu\alpha\alpha^{'}}c^{\dag}_{\alpha}c_{\alpha^{'}},\\
\hat{P}^{\dag}=\frac{1}{2}\sum_{\alpha}c^{\dag}_{\alpha}c^{\dag}_{\bar{\alpha}},\\
\hat{P}^{\dag}_{\mu}=\frac{1}{2}\sum_{\alpha,\alpha^{'}}Q_{\mu\alpha\alpha^{'}}c^{\dag}_{\alpha}c^{\dag}_{\bar{\alpha^{'}}}.\\
\end{array}
\label{QP}
\end{equation}
In the above equations, $Q_{\mu\alpha\alpha^{'}}$ is the matrix
element of the one-body quadrupole operator, namely
$\langle\alpha\arrowvert\hat{Q}_{2\mu}\arrowvert\alpha^{'}\rangle$
in which $\alpha$ represents the spherical single-particle state
denoted by $\{nljm\}$. $c^{\dag}_{\alpha}$ is the particle creation
operator on the corresponding state and its time reversal is defined
as $c_{\bar{\alpha}}=(-1)^{j-m}c_{nlj-m}$.

When the eigenvalue equation (Eq.~\eqref{eigeneq}) is solved, the
eigenstates can be determined. Correspondingly, the electric
quadrupole transition probabilities between the states
$\arrowvert\Psi^{I\sigma}\rangle$ and
$\arrowvert\Psi^{I^{'}\sigma^{'}}\rangle$ can be calculated by the
quadrupole operator (Eq.~\eqref{QP}):
\begin{equation}
B(E2,I\sigma\to
I^{'}\sigma^{'})=\frac{2I^{'}+1}{2I+1}\arrowvert\langle\Psi^{I^{'}\sigma^{'}}\Arrowvert\hat{Q}_{2}
\Arrowvert\Psi^{I\sigma}\rangle\arrowvert^{2}, \label{B(E2)}
\end{equation}
where the reduced matrix element is defined as
\begin{equation}
\begin{array}{c}
\langle\Psi^{I^{'}\sigma^{'}}\Arrowvert\hat{Q}_{2}
\Arrowvert\Psi^{I\sigma}\rangle=
$$\sum\limits_{{KK}^{'},{\kappa\kappa}^{'},\nu}(IK^{'}-\nu,2\nu\arrowvert I^{'}K^{'})\times$$ \\
\langle\Phi_{{\kappa}^{'}}\arrowvert\hat{Q}_{2\mu}\hat{P}^{I}_{K^{'}-\nu,K}\arrowvert\Phi_{\kappa}\rangle
F^{I^{'}\sigma^{'}}_{K^{'}\kappa^{'}}F^{I\sigma}_{K\kappa}.
\end{array}
\label{Belemt}
\end{equation}

\section{Results}

In the calculation, Nilsson's parameters ($\kappa$, $\mu$) for
$^{230,232}$Th, $^{232,234,236}$U and $^{240}$Pu are taken from
Refs.~\cite{Nil1969} and ~\cite{BR1985} and the shapes of the
Nilsson's deformed field for each nucleus are fixed. They are
described by $\epsilon_{2}$ and $\epsilon_{4}$ for quadrupole and
hexadecapole deformations which are listed in Tab.~\ref{Tab1}. The
$\epsilon_{2}$ value (Bear in mind that the relation between
$\epsilon_{2}$ and $\beta_{2}$ is approximately $\epsilon_{2}$ =
$\beta_{2}$$\times$0.95) is fixed for each nucleus changing from
0.212 to 0.260. We see from Tab.~\ref{Tab1} that the
$\varepsilon_2$ values of quadrupole deformations increase as the
numbers of valence nucleons increase. They are approximately in
accordance with the results of nonrelativistic mean-field
calculation with Gogny force~\cite{Delache} and the Relativistic
Mean-Field (RMF) calculation~\cite{Abusara}. Meanwhile, the
hexadecapole deformation parameter $\epsilon_{4}$ is nearly
one-order smaller than $\epsilon_{2}$.

\begin{table}
\centering
\caption{\label{Tab1} The quadrupole and hexadecapole deformation
parameters for $^{230,232}$Th, $^{232,234,236}$U and $^{240}$Pu,
respectively.}
\renewcommand\arraystretch{1.3}
\begin{tabular}{cccccccc}
\hline
\hline
 &$^{230}$Th&$^{232}$Th&$^{232}$U&$^{234}$U&$^{236}$U&$^{240}$Pu\\\hline
$\epsilon$$_{2}$&0.212&0.234&0.238&0.240&0.254&0.260\\
$\epsilon$$_{4}$&0.013&0.018&0.012&0.027&0.030&0.040\\
\hline\hline
\end{tabular}
\end{table}

The difference between the current HSM and PSM is that the
collective excitations described by $D_0$- and $D_2$-pair are
included in the basis space (see Eq.~\eqref{basis}). The first
thing we need to check is the collectivity of $D_0$- and
$D_2$-pair. In the single particle space (three major shells, N =
4, 5, 6 for protons and N = 5, 6, 7 for neutrons), the number of
$K$ =0 and $K$ = 2 2-qp states is about 60 and 80, respectively,
in the case of truncation energy 5 MeV. The main components
(percentages are larger than 2\%) of $D_0$- and $D_2$-pair are
listed in Tab.~\ref{Tab2} and Tab.~\ref{Tab3} for $^{232,234}$Th,
$^{232-236}$U and $^{240}$Pu, respectively.

In Tab.~\ref{Tab2}, we notice that for neutron configurations,
except the 2-qp state
$\frac{5}{2}^{+}[633]_{\nu}-\frac{5}{2}^{+}[622]_{\nu}$, all the
others are composed of one qp state and its time reversal partner.
The basis $\frac{1}{2}^{+}[631]_{\nu}-\frac{1}{2}^{+}[631]_{\nu}$
plays an important role for all the nuclei studied. On the other
hand, for all the nuclei except $^{240}$Pu, the basis
$\frac{1}{2}^{-}[501]_{\nu}-\frac{1}{2}^{-}[501]_{\nu}$ has very
large percentage. Except for $^{230}$Th, the configuration
$\frac{7}{2}^{-}[743]_{\nu}-\frac{7}{2}^{-}[743]_{\nu}$
has obvious distributions in the $D_{0}$-pairs.                                          
The percentages of
$\frac{1}{2}^{+}[631]_{\nu}-\frac{1}{2}^{+}[631]_{\nu}$ and
$\frac{5}{2}^{+}[622]_{\nu}-\frac{5}{2}^{+}[622]_{\nu}$ increase
as the neutron number increases due to the shift of the fermi
surface. The bases from the proton shell do not play such an
important role as those from the neutron shell.

\begin{table}
\centering \caption{ \label{Tab2}The main configurations of the
$D_{0}$-pairs constructed as in Eq.~\eqref{betagamma} for
$^{230,232}$Th, $^{232,234,236}$U and $^{240}$Pu, respectively.}
\renewcommand\arraystretch{1.3}
\begin{tabular}{c|c|c|c|c|c|c}
\hline\hline 2-qp
basis&$^{230}$Th&$^{232}$Th&$^{232}$U&$^{234}$U&$^{236}$U&$^{240}$Pu\\\hline
$\frac{5}{2}^{-}[503]_{\nu}-\frac{5}{2}^{-}[503]_{\nu}$&$<$2\%&$<$2\%&7.4\%&2.3\%&2.2\%&$<$2\%\\\hline
$\frac{1}{2}^{-}[501]_{\nu}-\frac{1}{2}^{-}[501]_{\nu}$&71.2\%&65.7\%&24.3\%&57.0\%&46.9\%&2.5\%\\\hline
$\frac{5}{2}^{+}[633]_{\nu}-\frac{5}{2}^{+}[622]_{\nu}$&$<$2\%&$<$2\%&2.3\%&$<$2\%&$<$2\%&$<$2\%\\\hline        
$\frac{13}{2}^{+}[606]_{\nu}-\frac{13}{2}^{+}[606]_{\nu}$&$<$2\%&$<$2\%&4.3\%&$<$2\%&$<$2\%&$<$2\%\\\hline      
$\frac{1}{2}^{+}[631]_{\nu}-\frac{1}{2}^{+}[631]_{\nu}$&4.7\%&10.2\%&8.4\%&14.5\%&22.2\%&26.6\%\\\hline         
$\frac{5}{2}^{+}[622]_{\nu}-\frac{5}{2}^{+}[622]_{\nu}$&$<$2\%&$<$2\%&$<$2\%&$<$2\%&8.6\%&48.7\%\\\hline         
$\frac{5}{2}^{-}[752]_{\nu}-\frac{5}{2}^{-}[752]_{\nu}$&6.1\%&$<$2\%&2.0\%&$<$2\%&$<$2\%&$<$2\%\\\hline
$\frac{7}{2}^{-}[743]_{\nu}-\frac{7}{2}^{-}[743]_{\nu}$&$<$2\%&10.2\%&37.3\%&11.3\%&7.3\%&10.6\%\\\hline\hline
\end{tabular}
\end{table}

The similar phenomena happen for the structure of $D_{2}$-pairs as
listed in Tab.~\ref{Tab3}. The 2-qp state
$\frac{3}{2}^{-}[501]_{\nu}+\frac{1}{2}^{-}[501]_{\nu}$ has about
25$\%$ percentages for both $^{230,232}$Th and $^{232,234,236}$U.
For $^{230,232}$Th and $^{232,234,236}$U, the configuration
$\frac{5}{2}^{-}[503]_{\nu}-\frac{1}{2}^{-}[501]_{\nu}$ plays a
very important role with the percentage about 50$\%$, but less
than 2$\%$ for $^{240}$Pu. The 2-qp configuration
$\frac{5}{2}^{+}[622]_{\nu}-\frac{1}{2}^{+}[631]_{\nu}$ has a
percentage of 98.3$\%$ for $^{240}$Pu and 11.6$\%$ for $^{236}$U,
but very small for the other nuclei. The structure of $D_2$ pairs
agrees well with the results in Ref.~\cite{Zuker65}, where the
structure of $\gamma$-vibrational states were investigated for
rare-earth and actinide-region nuclei by quasi-particle and
quasi-boson approximation. From Tabs.~\ref{Tab2} and ~\ref{Tab3},
we see that the $D$-pairs are composed of several 2-qp bases for
all the studied nuclei except $^{240}$Pu, indicating the
collectivity of $D$-pairs we constructed. Although there is only
one main component of 2-qp state in $D_{2}$-pairs for $^{240}$Pu,
it is a collective combination of several shell-model sp states.

\begin{table}
\centering \caption{\label{Tab3}The same as Tab.~\ref{Tab2}, but
for $D_{2}$-pairs.}
\renewcommand\arraystretch{1.3}
\begin{tabular}{c|c|c|c|c|c|c}
\hline\hline 2-qp
basis&$^{230}$Th&$^{232}$Th&$^{232}$U&$^{234}$U&$^{236}$U&$^{240}$Pu\\\hline
$\frac{3}{2}^{-}[501]_{\nu}+\frac{1}{2}^{-}[501]_{\nu}$&25.8\%&24.7\%&26.3\%&25.4\%&23.5\%&$<$2\%\\\hline
$\frac{5}{2}^{-}[503]_{\nu}-\frac{1}{2}^{-}[501]_{\nu}$&50.1\%&47.6\%&52.2\%&50.9\%&50.1\%&$<$2\%\\\hline
$\frac{3}{2}^{+}[631]_{\nu}+\frac{1}{2}^{+}[631]_{\nu}$&2.9\%&3.3\%&3.1\%&2.8\%&$<$2\%&$<$2\%\\\hline
$\frac{5}{2}^{+}[633]_{\nu}-\frac{1}{2}^{+}[631]_{\nu}$&5.4\%&7.4\%&7.1\%&6.3\%&2.6\%&$<$2\%\\\hline
$\frac{5}{2}^{+}[622]_{\nu}-\frac{1}{2}^{+}[631]_{\nu}$&$<$2\%&3.4\%&$<$2\%&2.5\%&11.6\%&98.3\%\\\hline
$\frac{7}{2}^{-}[743]_{\nu}-\frac{3}{2}^{-}[761]_{\nu}$&2.4\%&$<$2\%&$<$2\%&$<$2\%&$<$2\%&$<$2\%\\\hline
$\frac{3}{2}^{+}[402]_{\pi}+\frac{1}{2}^{+}[400]_{\pi}$&$<$2\%&2.5\%&$<$2\%&$<$2\%&2.1\%&$<$2\%\\\hline\hline
\end{tabular}
\end{table}

\begin{figure}
\centerline{\includegraphics[width=8.0cm,angle=0,clip]{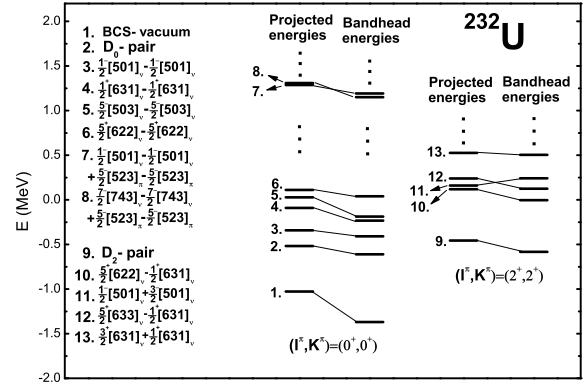}}
\caption{\label{Fig1} Comparison of the angular momentum projected
energies and the bandhead energies for ($I^\pi, K^\pi$)=(0$^+,
0^+$) and ($I^\pi, K^\pi$)=(2$^+, 2^+$) states of $^{232}$U,
respectively. The projected energies represent the values of
$H^{I=0}_{0\kappa,0\kappa}/N^{I=0}_{0\kappa,0\kappa}$ and
$H^{I=2}_{2\kappa,2\kappa}/N^{I=2}_{2\kappa,2\kappa}$,
respectively. The bandhead energies indicate the energies after
the diagonalization of Hamiltonian matrix (see
Eq.~\eqref{Helemt}). }
\end{figure}

When one solves Eq.~\eqref{eigeneq}, the angular momentum
projected energies are then mixed through the diagonalization of
the shell model Hamiltonian in Eq.~\eqref{Helemt}. The comparison
of the projected energies and the bandhead energies is helpful to
identify each band's configuration. As an example, in
Fig.~\ref{Fig1}, we plot the bandhead energies before and after
diagonalization for different projected states with ($I^\pi,
K^\pi$)=(0$^+, 0^+$) and ($I^\pi, K^\pi$)=(2$^+, 2^+$) of
$^{232}$U, respectively. We see from Fig.~\ref{Fig1} that both the
BCS vacuum state and the $D_{0}$-pair have very low energies,
while the latter one is about 500 keV higher. After the
diagonalization, the ground state become nearly 300 keV lower,
which indicates that to some extent the vacuum state mix with the
multi-qp states. And the similar phenomena happens for the other
$K^{\pi}$=0$^{+}$ states. For the $K^{\pi}$=2$^{+}$ states, the
diagonalization does not make a big difference as that of the
$K^{\pi}$=0$^{+}$ states does. In other words, the
$K^{\pi}$=2$^{+}$ states do not mix so much with each other. The
energies of $D_{0}$- and $D_{2}$-pairs before diagonalization have
very little difference with the $\beta$- and $\gamma$-bandhead
energies, respectively. Therefore it can be concluded the method
to construct the collective pairs as Eq.~\eqref{betagamma} is very
effective.

\begin{figure}
\centerline{\includegraphics[width=8.0cm,angle=0,clip]{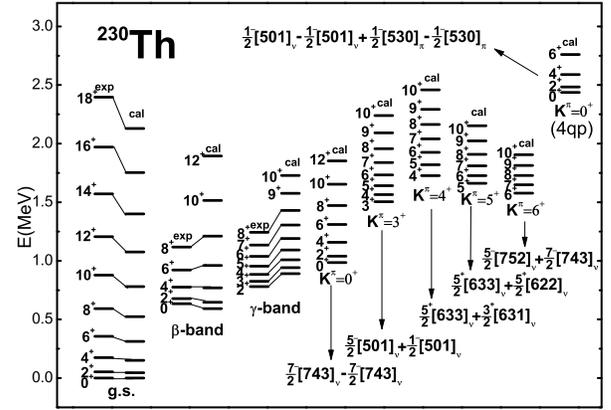}}
\caption{\label{Fig2} Comparison of the calculated and
experimental g-bands, $\beta$- and $\gamma$-band of $^{230}$Th.
Some 2-qp and 4-qp rotational bands are also given as a
theoretical prediction. The experimental energies are from the
National Nuclear Data Center~\cite{NNDC} and references therein.}
\end{figure}

Based on the collectivity of $D$-pairs, we obtain a more powerful
HSM by Extending the PSM basis with collective excitations, which
is a multi-shell model and valid for both qp excitations and
low-lying collective excitations such as $\beta$- and
$\gamma$-vibration. We solve the eigenvalue Eq.~\eqref{eigeneq} in
the basis space given by Eq.~\eqref{basis} and get the energy
levels and wavefunctions. Then the $B(E2)$ transitions are
calculated by the Eq.~\eqref{B(E2)}. As a first systemic numerical
realization of HSM, we calculate the $\beta$- and $\gamma$-bands,
some 2-qp and 4-qp rotational bands and the $B(E2)$ transition
rates for $^{230,232}$Th, $^{232,234,236}$U and $^{240}$Pu,
respectively.

For $^{230}$Th, we see from Fig.~\ref{Fig2} that the ground band
agrees well with the experimental data at low spins and has some
deviation at spins higher up to $I^{\pi}$=18$^{+}$. The agreement
between the $\beta$-band and $\gamma$-band with the corresponding
experimental values is also quite good. Our calculation predicts
five 2-qp rotational bands at 985 keV, 1504 keV, 1726 keV, 1661
keV and 1578 keV with $K^{\pi}$=0$^{+}$, $K^{\pi}$=3$^{+}$,
$K^{\pi}$=4$^{+}$, $K^{\pi}$=5$^{+}$ and $K^{\pi}$=6$^{+}$,
respectively. A $K^{\pi}$=0$^{+}$ 4-qp rotational band is given at
2437 keV with the configuration
$\frac{1}{2}^{-}[501]_{\nu}-\frac{1}{2}^{-}[501]_{\nu}+\frac{1}{2}^{-}[530]_{\pi}-\frac{1}{2}^{-}[530]_{\pi}$.

\begin{figure}
\centerline{\includegraphics[width=8.0cm,angle=0,clip]{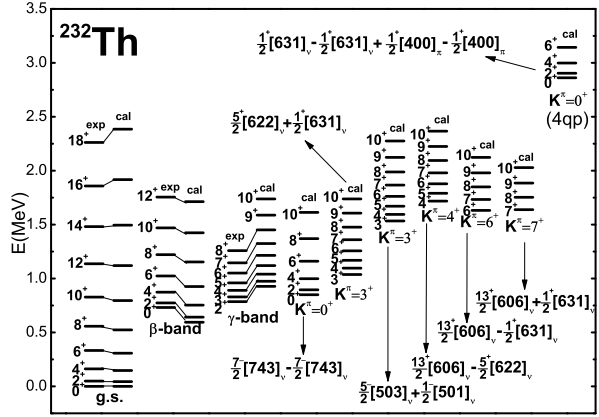}}
\caption{\label{Fig3} Same as Fig.~\ref{Fig2}, but for
$^{232}$Th.}
\end{figure}

In Fig.~\ref{Fig3}, we plot for $^{232}$Th several low-lying
multi-qp excited bands and collective bands from HSM and from some
available experiment data. We find that the ground band is in good
agreement with the experimental values up to spin
$I^{\pi}$=18$^{+}$. The calculated $\beta$-band is lower than the
observed one, obviously, while the case for $\gamma$-band is in
contrast. The not-very-good reproduction of the $\gamma$-band may
be due to the non-axial deformation and softness of the realistic
potential of this nucleus. It will be discussed at the end of this
section. Another $K^{\pi}=0^{+}$ band is predicted at 850 keV with
the configuration
$\frac{7}{2}^{-}[743]_{\nu}-\frac{7}{2}^{-}[743]_{\nu}$. Also,
there are two $K^{\pi}=3^{+}$ bands at 1037 keV and 1533 keV with
the configuration
$\frac{5}{2}^{+}[622]_{\nu}+\frac{1}{2}^{+}[631]_{\nu}$ and
$\frac{5}{2}^{-}[503]_{\nu}+\frac{1}{2}^{-}[501]_{\nu}$,
respectively. At 1718 keV, 1631 keV and 1640 keV, three bands with
$K^{\pi}=4^{+}$, $K^{\pi}=6^{+}$ and $K^{\pi}=7^{+}$ are
predicted, respectively. In our calculation, one 4-qp
$K^{\pi}=0^{+}$ rotational band is predicted at 2862 keV, with the
configuration
$\frac{1}{2}^{+}[631]_{\nu}-\frac{1}{2}^{+}[631]_{\nu}+\frac{1}{2}^{+}[400]_{\pi}-\frac{1}{2}^{+}[400]_{\pi}$.

\begin{figure}
\centerline{\includegraphics[width=8.0cm,angle=0,clip]{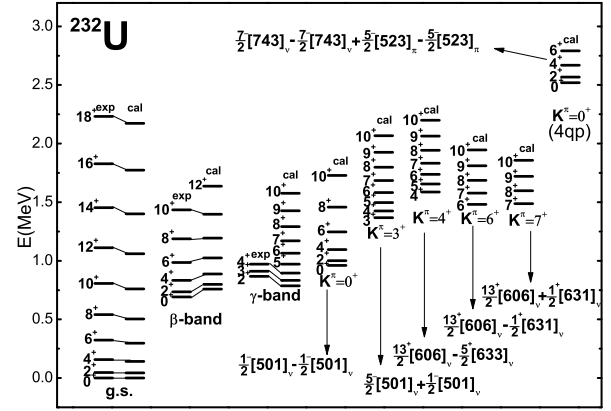}}
\caption{\label{Fig4} Same as Fig.~\ref{Fig2}, but for $^{232}$U.}
\end{figure}

The energy scheme of $^{232}$U is given in Fig.~\ref{Fig4}. We
find that the calculation well reproduces the ground band,
$\beta$- and $\gamma$-bands. According to the calculation, five
2-qp rotational bands emerge at 960 keV, 1367 keV, 1587 keV, 1481
keV and 1487 keV with $K^{\pi}$=0$^{+}$, $K^{\pi}$=3$^{+}$,
$K^{\pi}$=4$^{+}$, $K^{\pi}$=6$^{+}$ and $K^{\pi}$=7$^{+}$,
respectively. A low-lying 4-qp rotational band with
$K^{\pi}$=0$^{+}$ is predicted at 2520 keV with the configuration
$\frac{7}{2}^{-}[743]_{\nu}-\frac{7}{2}^{-}[743]_{\nu}+\frac{5}{2}^{-}[523]_{\pi}-\frac{5}{2}^{-}[523]_{\pi}$.

\begin{figure}
\centerline{\includegraphics[width=8.0cm,angle=0,clip]{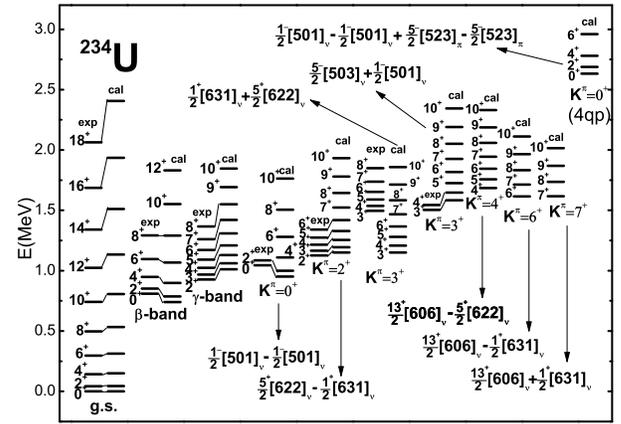}}
\caption{\label{Fig5} Same as Fig.~\ref{Fig2}, but for $^{234}$U.}
\end{figure}

The spectrum is shown in Fig.~\ref{Fig5} for $^{234}$U. For the
ground band, there are visible deviations between the observed
values and calculated ones when the spin is larger than 12$^{+}$,
but at low spin, the calculation agrees quite well with
experimental data. The calculated $\beta$- and $\gamma$-bands at
740 keV and 1012 keV have some differences, although not large,
with the experimental ones which are at 810 keV and 927 keV,
respectively, and moreover, the deviations become larger as the
spins increase. A $K^{\pi}$=0$^{+}$ band with the configuration
$\frac{1}{2}^{-}[501]_{\nu}-\frac{1}{2}^{-}[501]_{\nu}$ is given
at 952 keV, and the observed one is at 1044 keV. A $K^{\pi}=2^{+}$
2-qp band with the configuration
$\frac{5}{2}^{+}[622]_{\nu}-\frac{1}{2}^{+}[631]_{\nu}$ is given
at 1150 keV in our calculation, which is nearly the same as the
observed value 1125 keV. At 1136 keV and 1584 keV there are two
bands both with $K^{\pi}=3^{+}$ compared with two observed ones at
1496 keV and 1502 keV, respectively. The $K^{\pi}=6^{+}$ and
$K^{\pi}=7^{+}$ bands at 1611 keV and 1617 keV are also given as a
prediction. A $K^{\pi}=0^{+}$ 4-qp rotational band is given at
2633 keV with the configuration
$\frac{1}{2}^{-}[501]_{\nu}-\frac{1}{2}^{-}[501]_{\nu}+\frac{5}{2}^{-}[523]_{\nu}-\frac{5}{2}^{-}[523]_{\nu}$.

\begin{figure}
\includegraphics[width=8.0cm,angle=0,clip]{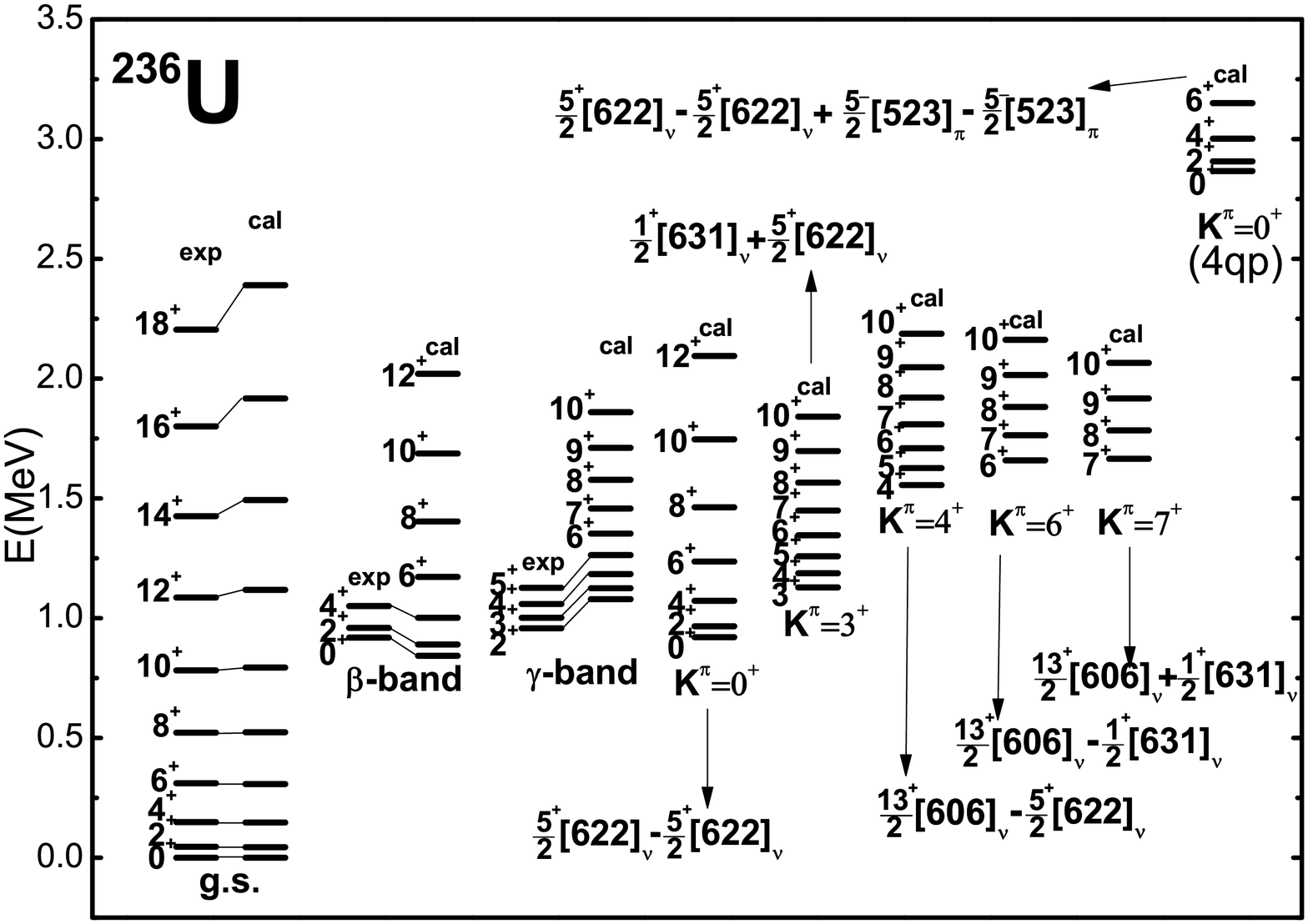}
\caption{\label{Fig6} Same as Fig.~\ref{Fig2}, but for $^{236}$U.}
\end{figure}

In Fig.~\ref{Fig6}, the energy scheme for $^{236}$U is plotted and
compared with the available experimental values. The calculated
ground band agrees well with the observed values up to spin
$I^{\pi}$=18$^{+}$. However, the calculated $\beta$- and
$\gamma$-bands have some deviations, although not large, from the
experimental values.
 A $K^{\pi}=0^{+}$ 2-qp band with configuration
$\frac{5}{2}^{+}[622]_{\nu}-\frac{5}{2}^{+}[622]_{\nu}$ is plotted
at 920 keV as a prediction. Another four 2-qp rotational bands are
given at 1129 keV, 1556 keV, 1658 keV and 1665 keV with
$K^{\pi}=3^{+}$, $K^{\pi}=4^{+}$, $K^{\pi}=6^{+}$ and
$K^{\pi}=7^{+}$, respectively. A $K^{\pi}=0^{+}$ 4-qp rotational
band is given at 2867 keV, with the configuration
$\frac{5}{2}^{+}[622]_{\nu}-\frac{5}{2}^{+}[622]_{\nu}+\frac{5}{2}^{-}[523]_{\pi}-\frac{5}{2}^{-}[523]_{\pi}$.

 \begin{figure}
\includegraphics[width=8.0cm,angle=0,clip]{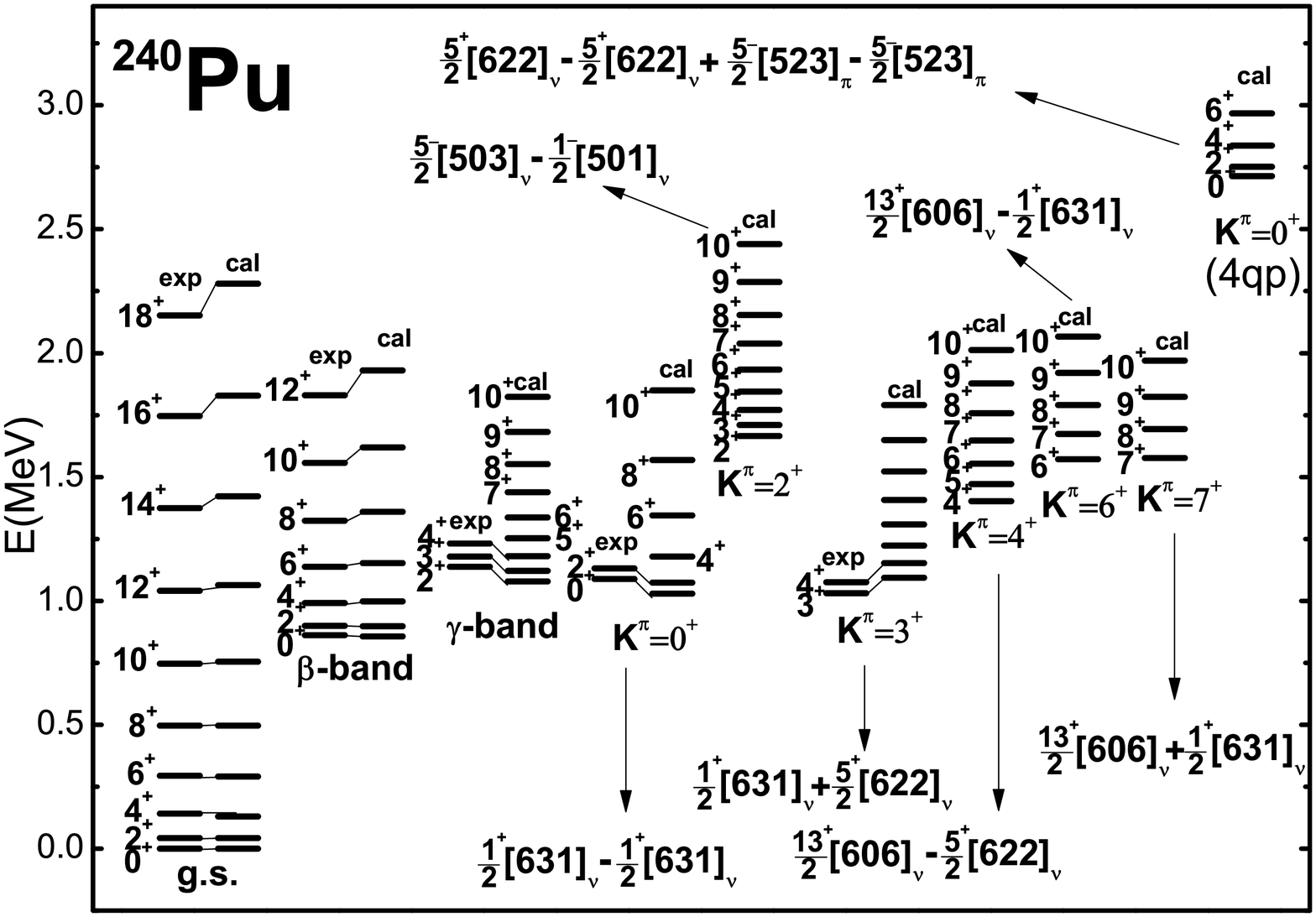}
\caption{\label{Fig7} Same as Fig.~\ref{Fig2}, but for
$^{240}$Pu.}
\end{figure}

The energy scheme of $^{240}$Pu is shown in Fig.~\ref{Fig7}. Both
the calculated ground band, $\beta$- and $\gamma$-band and $K=3^+$
band agree well with the experimental data. In the calculation,
the most important configuration of the $\gamma$-band is
$\frac{5}{2}^{+}[622]_{\nu}-\frac{1}{2}^{+}[631]_{\nu}$ which is
shown in the structure of $D_{2}$-pair in Tab.~\ref{Tab3} and that
makes almost no difference  with the results of
Ref.~\cite{Zuker65}. The calculation also well reproduces 2-qp
rotational bands with $K^{\pi}$=0$^{+}$ and $K^{\pi}$=3$^{+}$ at
1029 keV and 1094 keV compared to the experimental values 1089 keV
and 1031 keV, respectively. Moreover, the configuration of
$K^{\pi}$=3$^{+}$ band is
$\frac{5}{2}^{+}[622]_{\nu}+\frac{1}{2}^{+}[631]_{\nu}$ in our
calculation, which is the same as that suggested in
Ref.~\cite{NNDC}. Another calculated $K^\pi=2^{+}$ band with the
configuration
$\frac{5}{2}^{-}[503]_{\nu}-\frac{1}{2}^{-}[501]_{\nu}$ is
predicted with the bandhead energy 1666 keV.

In Ref.~\cite{LZP10}, $^{240}$Pu is studied in the framework of
the three-dimensional relativistic Hartree-Bogoliubov calculation
with the density-dependent, point-coupling energy density
functional, and in the $\beta$-$\gamma$ plane, the minimum of
binding energy is at the point with $\gamma$=0$^\circ$ and
$\beta_{2}$=0.280, which indicates the axial symmetry shape. The
current HSM is constructed under the assumption of axial symmetry,
and the $\epsilon_{2}$ (0.260) we choose is very close to the
shape suggested in Ref.~\cite{LZP10}.

\begin{widetext}
\begin{table*}
\centering \caption{\label{Tab4}Comparison of the $B(E2)$ values (in
the unit of W.u.) between the calculated results and the
experimental data. The experimental data are from
Ref.~\cite{NNDC}}.

\renewcommand{\arraystretch}{1.3}
\begin{tabular}{c|cc|cc|cc|cc|cc|cc}
\hline\hline
$B(E2)$&\multicolumn{2}{c|}{$^{230}$Th}&\multicolumn{2}{c|}{$^{232}$Th}&\multicolumn{2}{c|}{$^{232}$U}&\multicolumn{2}{c|}{$^{234}$U}&\multicolumn{2}{c|}{$^{236}$U}&\multicolumn{2}{c}{$^{240}$Pu}\\\hline
$I_{i}$$\to$$I_{f}$&exp&cal&exp&cal&exp&cal&exp&cal&exp&cal&exp&cal\\\hline
4$_{g}^{+}$$\to$2$_{g}^{+}$&265(9)&289.7&286(24)&347.8&\textbf{---}&376.5&\textbf{---}&379.6&357(23)&409.3&\textbf{---}&439.6\\\hline
2$_{g}^{+}$$\to$0$_{g}^{+}$&196(6)&202.0&198(11)&242.7&241(21)&263.1&236(10)&265.0&250(10)&285.8&287(11)&307.3\\\hline
2$_{\beta}^{+}$$\to$0$_{g}^{+}$&2.7(9)&0.37&2.8(12)&0.55&\textbf{---}&0.64&$<$1.3&0.54&\textbf{---}&0.45&\textbf{---}&0.10\\\hline
2$_{\gamma}^{+}$$\to$0$_{g}^{+}$&2.9(9)&1.27&2.9(4)&1.16&\textbf{---}&1.32&2.9(5)&1.28&\textbf{---}&0.56&\textbf{---}&0.02\\\hline
2$_{\gamma}^{+}$$\to$0$_{\beta}^{+}$&\textbf{---}&0.10&\textbf{---}&0.04&\textbf{---}&1.1&\textbf{---}&0.01&\textbf{---}&0.04&\textbf{---}&0.05\\\hline\hline
\end{tabular}
\end{table*}
\end{widetext}

When the wavefunctions of the initial and final states are gotten,
we calculate the reduced $B(E2)$ transition probabilities between
them according to Eq.~(\ref{B(E2)}). The inter-band $B(E2)$ value
is a quantity that indicates the $K$ mixing in different bands. In
Tab.~\ref{Tab4}, the calculated intra-band $B(E2)$ values of
ground bands and inter-band ones from $\beta$-bands,
$\gamma$-bands to ground states are listed and compared with the
available observed values in Wisskoff unit (W.u.), respectively.
For the nuclei $^{230}$Th, $^{232}$U and $^{240}$Pu, the
calculated $B(E2)$'s from 2$^{+}_{g}$ to 0$^{+}_{g}$ agree well
with the experimental values, while for the other nuclei there
exists some difference between the calculation and experimental
data, especially for $^{232}$Th. For $^{230,232}$Th and $^{236}$U,
the calculated $B(E2)$ transitions from 4$^{+}_{g}$ to 2$^{+}_{g}$
can not reproduce the experimental data very well. The inter-band
transition probabilities are very small, nearly forbidden. For
example, for $^{230,232}$Th and $^{234}$U the 2$^{+}_{\beta}$ to
0$^{+}_{g}$ values are 0.37, 0.55 and 0.54 compared to the
experimental ones, 2.7, 2.8 and 1.3, respectively. Furthermore,
for these three nuclei, the experimental $B(E2)$'s from
2$^{+}_{\gamma}$ to 0$^{+}_{g}$ are all 2.9, but the calculated
ones are just 1.27, 1.16 and 1.28, respectively. Therefore, on the
whole, the calculated inter-band transition probabilities are
smaller than the experimental data for the transitions from
 $\beta$- or $\gamma$-band to the ground state. It indicates
 that in realistic nuclei, the potentials in both $\beta$ and $\gamma$
direction are stiffer than those assumed in the HSM, according to
the discussion in Ref.~\cite{Pie04,Cast}. The calculated $B(E2)$
transitions from 2$^{+}_{\gamma}$ to 0$^{+}_{\beta}$ are also very
small .

In the case of SU(3) limit, according to the FDSM or IBM, the
ground-state and the degenerated $\beta$- and $\gamma$-vibrational
states belong to different irreducible representations (irrps) of
the SU(3) group. The $\beta$- and $\gamma$-bands are distinguished
by different $K$ values, which means in this case, $B(E2)$
transitions between the inter-bands are forbidden. However, both
the calculated inter-bands $B(E2)$'s and experimental data are
non-zero, which indicates the mixing of the spaces with different
irrps.

\begin{figure}
\centerline{\includegraphics[width=8.0cm,angle=0,clip]{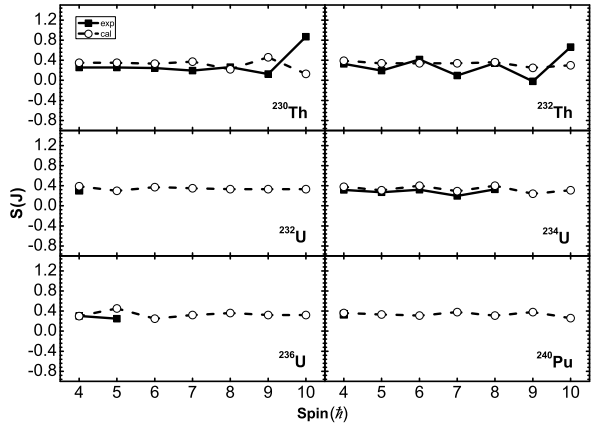}}
\caption{\label{Fig8} Comparison of the calculated $S(J)$ values
of $\gamma$-vibrational bands with experimental data for
$^{230,232}$Th, $^{232,234,236}$U and $^{240}$Pu, respectively. }
\end{figure}

In Refs.~\cite{Zam91,Stag07}, the benchmark
\begin{equation}
S(J)=\frac{\{E[J^{+}_{\gamma}]-E[(J-1)^{+}_{\gamma}]\}-\{E[(J-1)^{+}_{\gamma}]-
E[(J-2)^{+}_{\gamma}]\}}{E[2^{+}_{g}]},
\end{equation}
is defined to estimate the non-axiality and softness of the
$\gamma$ deformation. In the equation above, $E[J_{\gamma}^{+}]$
is the energy level of $\gamma$-bands with spin $J$, and
$E[2_{g}^{+}]$ is the energy of the first excited state of the
ground band. In the case of axially symmetric rotor, $S(J)$ is
equal to 0.333, and the staggering around this value indicates the
non-axial effect. In the microscopic viewpoint, the staggering
indicates the mixing of bases with different $K^{\pi}$s~
\cite{LZP10}. In Fig.~\ref{Fig8}, we plot the $S(J)$ values of
both the observed and calculated $\gamma$-bands as a function of
spin for all the six nuclei we studied. For $^{230}$Th, the
calculated $S(J)$ values have small deviations from experimental
data except at $J^{\pi}$=9$^{+}$ and 10$^{+}$. For $^{234}$U, the
observed and calculated $S(J)$ values are nearly the same, and
moreover, the staggering is still small. According to our
calculation, the $S(J)$'s for $^{232}$U and $^{240}$Pu nearly keep
constant 0.333 at low spins, well reproducing one experimental
data, respectively. The calculated $S(J)$'s nearly keep constant
at low spins for $^{232}$Th. However, the staggering of
experimental $S(J)$ data is obvious, indicating the non-axial
shapes of $^{232}$Th, and it may explain why the HSM calculation
does not well reproduce the experimental $\gamma$-band for this
nucleus. Moreover, for $^{236}$U, the staggering of the calculated
values is very small, which indicates a good axial shape.

\section{Conclusion}

In order to describe simultaneously the single-particle and
low-lying collective excitations for heavy nuclei, the PSM is
extended to the HSM by adding the collective degrees of freedom,
namely the $D$-pairs excitations, into PSM intrinsic basis. The
study about the structure of the $D$-pairs indicates the method to
construct $D_0$- and $D_2$-pair is reasonable by the linear
combination of all the 2-qp states with $K^{\pi}=0^{+}$
$(K^{\pi}=2^{+})$ in the PSM truncated space. In this way, the
$D_{0}$- and $D_2$-pair do show collectivity.

Based on the collectivity of $D$-pairs, the energy levels and
$B(E2)$ transitions for the g-band, 2-qp and 4-qp excitations, and
collective $\beta$-bands and $\gamma$-bands are described
simultaneously in HSM for deformed actinide nuclei $^{230,232}$Th,
$^{232,234,236}$U and $^{240}$Pu, respectively. The calculation
well reproduces the $g$-bands, $\beta$ and $\gamma$-bands and some
quasiparticle bands compared with the observed values, although
for $^{232}$Th, the deviations between the calculated and observed
$\gamma$-bands is big due to the non-axial deformations. In
addition, some low-lying quasiparticle bands are predicted,
awaiting experimental confirmation. For all the nuclei studied,
the calculated $B(E2)$ values in the g-bands from $2^{+}_{g}$ to
$0^{+}_{g}$ and from $4^{+}_{g}$ to $2^{+}_{g}$ and the inter-band
ones agree with the experimental values.

We demonstrate that the HSM can describe simultaneously low-lying
collective and quasi-particle excitations in deformed nuclei by
the collective 1-$D$-pairs. Meanwhile, the model space is still
kept tractable for heavy nuclear systems. Furthermore, HSM can
also study 2-phonon excitations by adding 2-$D$-pairs into the
intrinsic basis of PSM, which will be our future work. Along this
line, HSM will become a powerful multi-major-shell shell model,
useful for both well-deformed nuclei and transitional ones.

\section*{Acknowledgment}

Useful discussions with Y.-S. Chen are gratefully acknowledged.
This work was supported by National Natural Science Foundation of
China (Nos. 10975116, 11275160 and 11175258).

\end{document}